\documentclass[twocolumn]{revtex4}
\usepackage{graphicx}
\usepackage{bm}
\usepackage{color}
\usepackage{amsmath}
\usepackage{natbib}

\begin{document}

\title{On the rich eight branch spectrum of the oblique propagating longitudinal waves in partially spin polarized electron-positron-ion plasmas}

\author{Pavel A. Andreev}
\email{andreevpa@physics.msu.ru}
\affiliation{Faculty of physics, Lomonosov Moscow State University, Moscow, Russian Federation.}
\author{Z. Iqbal}%
\email{abbasiravian@yahoo.com}
\affiliation{Department of Physics, G. C. University Lahore, Katchery Road, Lahore 54000, Pakistan
and Faculty of physics, Lomonosov Moscow State University, Moscow, Russian Federation.}

\date{\today}

\begin{abstract}
 We consider the separate spin evolution of electrons and positrons in electron-positron and electron-positron-ion plasmas. We consider oblique propagating longitudinal waves in this systems. Working in a regime of high density $n_{0}\sim10^{27}$ cm$^{-3}$ and high magnetic field $B_{0}=10^{10}$ G we report presence of the spin-electron acoustic waves and their dispersion dependencies. In electron-positron plasmas, similarly to the electron-ion plasmas, we find one spin-electron acoustic wave (SEAW) at propagation parallel or perpendicular to the external field and two spin-electron acoustic waves at the oblique propagation. At the parallel or perpendicular propagation of the longitudinal waves in electron-positron-ion plasmas we find four branches: the Langmuir wave, the positron-acoustic wave and pair of waves having spin nature, they are the SEAW and, as we called it, spin-electron-positron acoustic wave (SEPAW). At the oblique propagation we find eight longitudinal waves: the Langmuir wave, Trivelpiece–-Gould wave, pair of positron-acoustic waves, pair of SEAWs, and pair of SEPAWs. Thus, for the first time, we report existence of the second positron-acoustic wave existing at the oblique propagation and existence of SEPAWs.
\end{abstract}

%\pacs{52.27.Ep   Electron-positron plasmas; 73.22.Lp Collective excitations;
%52.30.Ex Two-fluid and multi-fluid plasmas;
%52.35.Dm Sound waves}% PACS, the Physics and Astronomy
                             % Classification Scheme.
%\keywords{acoustic waves, electron-positron plasmas, quantum plasmas, quantum hydrodynamics, spin evolution}
%Use showkeys class option if keyword

\maketitle
%52.27.Ep   Electron-positron plasmas
%73.22.Lp	Collective excitations
%52.30.Ex	Two-fluid and multi-fluid plasmas
%52.35.Dm	Sound waves

%%%%%%%%%%TEXT

\section{\label{sec:level1} Introduction}

The field of spin quantum plasmas has been rapidly growing over the last
decade. Takabayasi \cite{Takabayasi1}
derived and analyzed the quantum hydrodynamic equations
for a single spin-1/2 particle. The effects of electron spin on the
plasma dynamics were first studied by Kuz'menkov at. al, \cite{kuzmenkov-
page-136-2001, kuzmenkov- page-258-2001} in 2001. These authors have
developed a method of explicit derivation of many-particle quantum hydrodynamic (QHD) equations. These equations were truncated to consist of
the continuity equation, the Euler equation,
the energy balance equation, the magnetic moment evolution equation for spin-1/2 quantum plasmas. The starting point of these
derivation was the many-particle Pauli equation. These set of equations
contain the effects of the spin-spin exchange interactions and Coulomb
exchange interactions. Another form of derivation was recently suggested in Ref. \cite{Koide PRC 13}. A simplified form of QHD equations
were considered in Refs. \cite{Marklund and Brodin-Phys. Rev. Lett.
98 (2007), Brodin and Marklund- New J. Phys. 9(2007)}.

The method of many particle QHDs was applied to
study the eigenwave problem for spin-1/2 electron-ion plasmas with an account of the ion motion \cite{Moscow University
Physics Bulletin-2007}.
The dispersion relations of electrostatic and electromagnetic waves has been
studied and found that the spin-plasma waves (found for electrons in Ref. \cite{Vagin BRAS 06}) exist in the vicinity of
electron and ion cyclotron frequencies in the spectrum of waves propagating
perpendicular to the external magnetic field, and the
dispersion relation for self-consistent spin waves with a linear spectrum is also obtained in Ref. \cite{Moscow University
Physics Bulletin-2007}. Further applications of the method of many particle
QHDs have been given in Refs. \cite{P. A. Phys.
Atomic Nuclei-2008, Int. J. Mod. Phys. B-2012} which predicts a mechanism of instabilities
which arises due to the interaction of neutron beam with electron-ion
magnetized spin-1/2 quantum plasma, including instability of the spin-plasma waves. Further analysis of spin-plasma waves was presented in Refs. \cite{Browdn-PRL 101- (2008), Misra-j-plasma phys-2010}. Considering kinetics in the extended phase space suggested by Kagan in 1961 \cite{Kagan JETP 61 a}-\cite{Kagan JETP 66}, where the spin or magnetic moment is considered along with the coordinate and momentum, the fine structure of the Berstein modes is found \cite{Hussain PP 14 spin bernst}. The fine structure is demonstrated on the example of the second mode. It arises due to the presence of the anomalous magnetic moment of electrons.

Recently, a set of QHD equations for
charged spin-1/2 particles is derived from the Pauli equation in Ref. \cite{Andreev PRE 15}. It forms separate spin evolution QHDs (SSE-QHDs) Which treats spin-up and spin-down electrons as two different
fluids. It is essential if the populations of spin-up electrons and spin-down
electrons in the presence of external magnetic field is different ($n_{\uparrow }\neq n_{\downarrow }$). This difference of populations of
quantum states is responsible for difference of Fermi pressures of the
spin-up and spin-down electrons. Revealing in a new type of a
soundlike solution called the spin-electron acoustic wave (SEAW) \cite{Andreev PRE 15}, see some additional discussion below, after formula (\ref{SEPAWOP Euler eq electrons spin UP}). Spin current evolution in terms of SSE-QHD was considered in Ref. \cite{Trukhanova PLA 15}.

In the framework of
the hydrodynamic model and linear-response-function formalism the effects of spin polarization on the Langmuir and zero sound waves is investigated in Ref. \cite{Margulis-Zh-
Eksp- Teor. Fiz- 9-1983}. It was found that spin polarization increases the coefficient of spatial dispersion of Langmuir waves. It has also shown
that phase velocity of zero sound increases with increase of the degree of
polarization.

Electron-positron (e-p) plasma is distinct because it consists of particles
which have mass symmetry and anti charge symmetry. Naturally
electron-positron plasmas are found in many astrophysical environments like
early universe \cite{Rees 1983}, in neutron star magnetosphere \cite{Paul-astro-phys-2007}, \cite{sturo}. Naturally
existence of electron-positron plasma in compact stars has been investigated
by applying a simplified model of a gravitationally collapsing or pulsating
baryon core. It has shown that possible electric processes that lead to the
production of electron-positron pairs in the boundary of a baryon core and
calculate the number density of electron-positron pairs $n_{pair}=10^{28}$ $cm^{-3}$ \cite{Han PRD 12}. The degenerate electron-positron plasmas with ions are believed
to be found in compact astrophysical bodies like neutron stars and the inner
layers of white dwarfs \cite{Lai-Rev. Mod. Phys. 73-2001, Harding RPP 06,
Shapiro-Germany-2004}.

Physicists also trying to generate e--p plasmas in
laboratories. In this context different schemes have been proposed for the
laboratory generation. For example, in large-scale
conventional accelerators, the possibility of recombining high-quality
electron and positron beams via magnetic chicanes \cite{Greaves-phys
plasma-1994}. Pederson et. al, \cite{Pederson-New J. Phys-2012} have been
presented plan for the creation and diagnosis of electron-positron plasmas
in a stellarator, based on extrapolation of the results from the Columbia
Non-neutral Torus stellarator, as well as recent developments in positron
sources. Interaction of ultrashort laser pulses with gaseous or solid
targets could lead to the generation of the optically thin e-p plasma with
above solid state densities in the range of ($10^{23}$ $-10^{28}$) $cm^{-3}$%
\cite{Eliasson - 2013}. Recently, it has shown that, by using a
compact laser-driven setup, ion-free electron--positron plasmas can be
generated in the laboratory \cite{Sarri}. Their charge neutrality, density about $10^{16}cm^{-3}$ and small
divergence finally open up the possibility of studying electron--positron
plasmas in controlled laboratory experiments.

The wave propagation phenomenon in electron-positron plasma is different as
in usual electron-ion plasma. Using a two-fluid model and a kinetic model,
it has been observed that many wave phenomena like acoustic waves, whistler
waves, Faraday rotation, lower hybrid waves and shear Alfven waves are
absent in the nonrelativistic e-p plasmas \cite{zank, Iwamoto}. Most of
applications related to the plasma wave phenomenon presented in above
mentioned studies has focused on the electron--ion spin quantum plasmas.
However, some applications were presented for electron-positron plasmas and
electron-positron-ion plasmas for both spin-1/2 quantum plasma and spinless
quantum plasma. For instance, the set of spin-1/2 QHD equations developed for electron-ion plasmas was applied for e-p plasmas by Brodin and Marklund \cite{Brodin and
Marklund-pop-2007}. They found new spin depended Alfv\'{e}nic solitary structures obeying the Korteweg–de Vries equation, where the nonlinearity is caused by spin effects. Mushtaq et. al \cite{Mushtaq-pop-2012}, \cite{Maroof-pop-2015} studied the effects of quantum Bohm potential and spin corrections on the spectrum of
magnetosonic waves in non- relativistic and relativistic degenerate electron-positron-ion (e-p-i)
plasmas, where the relativistic effects are included in the Fermi pressure only.
A hydrodynamic and kinetic models for spin-1/2 electron-positron quantum
plasmas has been developed in Ref. \cite{PA Phys. Plasmas-22-062113(2015)}
which incorporates the Coulomb, spin-spin, Darwin and annihilation
interactions. There was concluded that
the contributions of the annihilation interactions shifts the
eigen-frequencies of the transverse electromagnetic plane polarized waves
and transverse spin-plasma waves.

New longitudinal wave in the degenerate e-p-i
spinless quantum plasma has been reported in Refs. \cite{Nejoh-Aust-j-phy-1996}--\cite{Tribeche PP 09}.
It was called positron acoustic wave (PAW). For ultrarelativistic electrons and
non-relativistic positrons the dispersion relation of PAWs in the intermediate wave range were obtained \cite{Tsintsadze-j.plasma phy-2013}. The nonlinear wave
structure of large amplitude PAWs in e-p plasma with
electron beam has been discussed in Ref. \cite{Nejoh-Aust-j-phy-1996}.

In the present work, we employ the separated spin evolution QHD for the
e-p and e-p-i magnetized degenerate plasmas.
At consideration of separate evolution of spin-up and spin-down
electrons and positrons we discuss the oblique propagation of
longitudinal waves. So, in the present work we calculate the
spectrum of SEAWs, PAWs and predict spin-electron-positron acoustic waves.

\section{\label{sec:level1}Analytical model}

The SSE-QHDs developed in Refs. \cite{Andreev PRE 15}, \cite{Andreev AoP 15} can be applied to the electron-positron plasmas and electron-positron-ion plasmas. Therefore, we present the continuity and Euler equations for each spin projection of each species.

In this paper we consider the evolution of the longitudinal waves in non-relativistic plasmas. Hence the spin evolution does not affects our results. The Fermi spin current (the thermal part of the spin current for the degenerate spin-1/2 fermions) obtained in Ref. \cite{Andreev 1510 Spin Current} is not considered here. Explicit spin contribution in the spectrum of the Langmuir waves arising via the spin-orbit interaction is found in Ref. \cite{Ivanov PTEP 15} (see formula (34) for describing propagation perpendicular to the external magnetic field). In our approximation we need the continuity and Euler equations for each subspecies.

The continuity equation in the SSE-QHD arises as follows \cite{Andreev PRE 15}
\begin{equation}\label{SEPAWOP cont eq electrons spin s}
\partial_{t}n_{as}+\nabla(n_{as}\textbf{v}_{as})=(-1)^{i_{s}}T_{az}, \end{equation}
where $a=e,p$ for electrons and positrons correspondingly, $s=u, d$ for the spin-up and spin-down conditions of particles, $n_{as}$ and $\textbf{v}_{as}$ are the concentration and velocity field of particles of species $a$ being in the spin state $s$, $T_{az}=\frac{\gamma}{\hbar}(B_{x}S_{ay}-B_{y}S_{ax})$ is the z-projection of spin torque, $i_{s}$: $i_{u}=2$, $i_{d}=1$, with the spin density projections $S_{ax}$ and $S_{ay}$, each of them simultaneously describe evolution of the spin-up and spin-down particles of each species. Therefore, functions $S_{ax}$ and $S_{ay}$ do not bear subindexes $u$ and $d$. In this model the z-projection of the spin density $S_{az}$ is not an independent variable, it is a combination of concentrations $S_{az}=n_{au}-n_{ad}$.

The time evolution of the velocity fields of all species of particles for each projection of spin $\textbf{v}_{au}$ and $\textbf{v}_{ad}$ is governed by the Euler equations \cite{Andreev PRE 15}
$$mn_{as}(\partial_{t}+\textbf{v}_{as}\nabla)\textbf{v}_{as}+\nabla P_{as}$$
$$=q_{a}n_{as}\biggl(\textbf{E}+\frac{1}{c}[\textbf{v}_{as},\textbf{B}]\biggr)+(-1)^{i_{s}}\gamma_{a}n_{as}\nabla B_{z}$$
\begin{equation}\label{SEPAWOP Euler eq electrons spin UP} +\frac{\gamma_{a}}{2}(S_{ax}\nabla B_{x}+S_{ay}\nabla B_{y})+(-1)^{i_{s}}m(\widetilde{\textbf{T}}_{az}-\textbf{v}_{as}T_{az}),\end{equation}
with $P_{as}=(6\pi^{2})^{\frac{2}{3}}n_{as}^{\frac{5}{3}}\hbar^{2}/5m$, $\widetilde{\textbf{T}}_{az}=\frac{\gamma_{a}}{\hbar}(\textbf{J}_{(M)ax}B_{y}-\textbf{J}_{(M)ay}B_{x})$, which is the torque current, where
$\textbf{J}_{(M)ax}=(\textbf{v}_{au}+\textbf{v}_{ad})S_{ax}/2$, and $ \textbf{J}_{(M)ay}= (\textbf{v}_{au}+\textbf{v}_{ad})S_{ay}/2$ are the convective parts of the spin current tensor.  All species affect each other via the electric field: $\nabla \textbf{E}=4\pi\sum_{a,s}q_{as}n_{as}$ and $\nabla\times \textbf{E}=0$.

\begin{figure}
\includegraphics[width=8cm,angle=0]{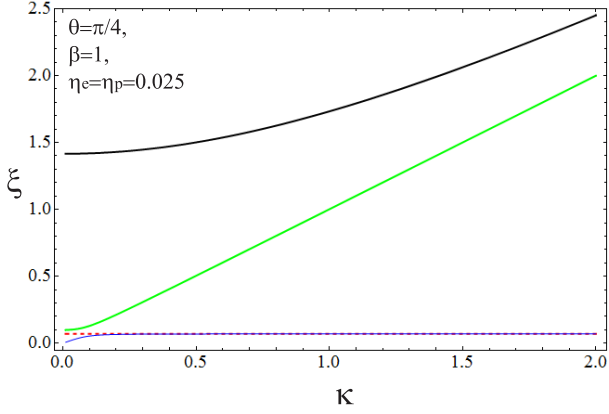}
\caption{\label{SEPAWOP F1epOb} (Color online) The figure shows the dispersion of the oblique propagating longitudinal waves in the electron-positron plasmas. It shows four waves. Upper branch describes the Langmuir wave. Second line from the bottom, which looks almost horizontal and depicted by the dashed line, presents the Trivelpiece–-Gould wave. Two other waves are the lower and upper branches of the spin-electron acoustic waves. Details of the low frequency branches are shown in the next figure.}
\end{figure}

\begin{figure}
\includegraphics[width=8cm,angle=0]{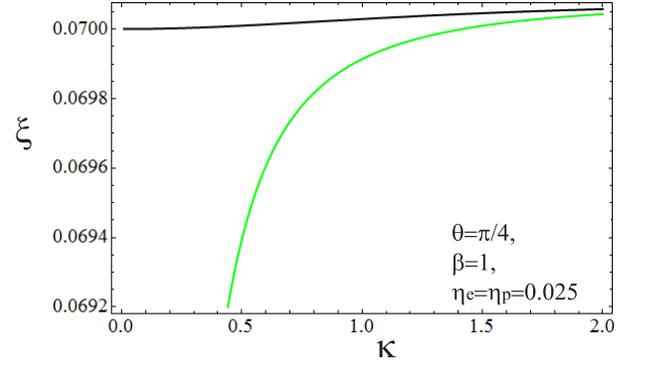}
\caption{\label{SEPAWOP F1epObSm} (Color online) The figure shows the low frequency part of spectrum of oblique propagating longitudinal waves in e-p plasmas, where $\Sigma=0.1$.}
\end{figure}

\begin{figure}
\includegraphics[width=8cm,angle=0]{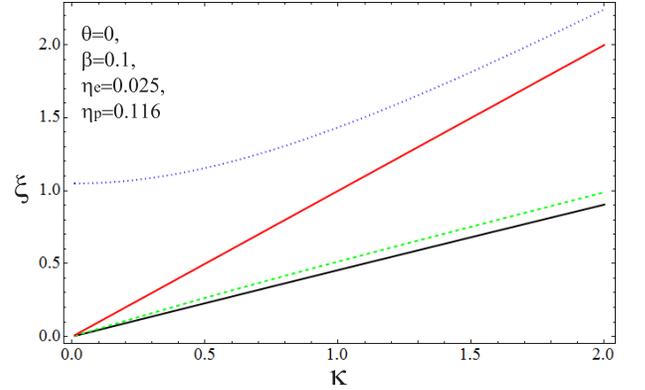}
\caption{\label{SEPAWOP F2} (Color online) The figure shows the longitudinal waves in e-p-i plasmas propagating parallel to the external magnetic field. Ratio between the electron and positron concentrations is chosen to be $\beta=0.1$. The upper branch shows the Langmuir wave dispersion. Three linear dependencies present SEAW, PAW and SEPAW.}
\end{figure}

\begin{figure}
\includegraphics[width=8cm,angle=0]{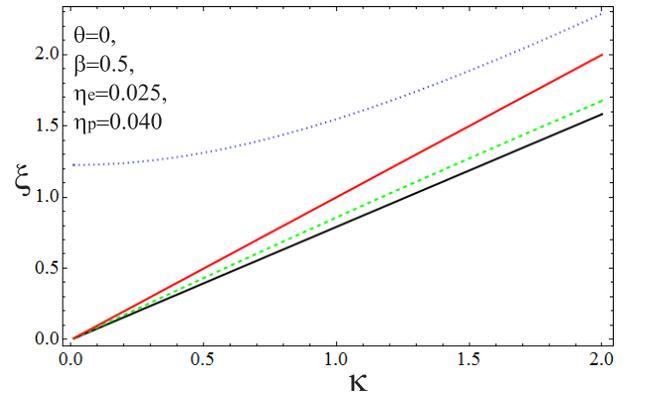}
\caption{\label{SEPAWOP F3} (Color online) The figure shows the longitudinal waves in e-p-i plasmas propagating parallel to the external magnetic field for $\beta=0.5$ and $\Sigma=0.1$.}
\end{figure}

\begin{figure}
\includegraphics[width=8cm,angle=0]{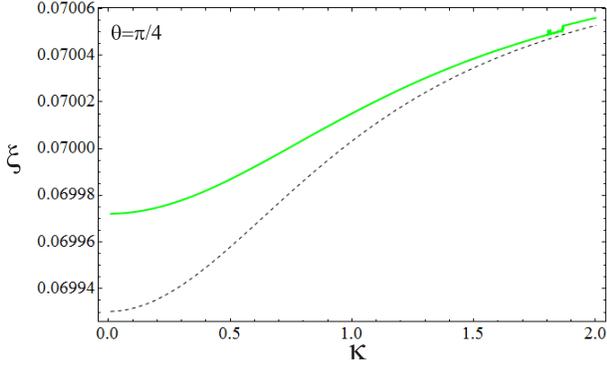}
\caption{\label{SEPAWOP F8} (Color online) The figure shows the Trivelpiece–Gould wave in the e-p-i plasmas for the different ratios between concentrations of electrons and positrons. Upper (lower) line is constructed for $\beta=0.5$ ($\beta=0.1$) and $\Sigma=0.1$.}
\end{figure}

\begin{figure}
\includegraphics[width=8cm,angle=0]{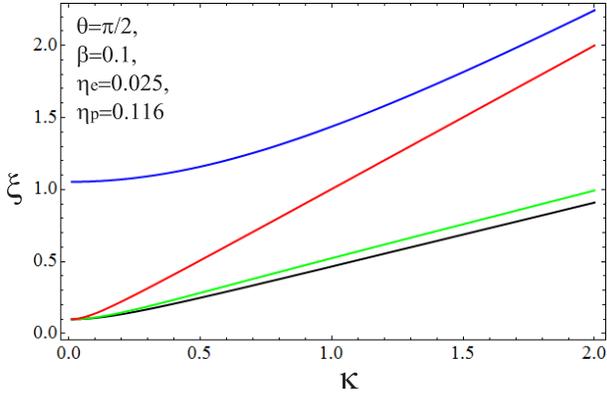}
\caption{\label{SEPAWOP F4} (Color online) The figure shows the longitudinal waves in e-p-i plasmas propagating perpendicular to the external magnetic field for $\beta=0.1$. As in Fig. \ref{SEPAWOP F2} we have four branches: Langmuir wave (in other words the upper hybrid wave), SEAW, PAW and SEPAW. It shows the frequency square shift for all branches on $\Omega^{2}$, with the dimensionless cyclotron frequency $\Sigma=0.1$.}
\end{figure}

The SSE-QHDs was applied
to two-dimensional electron gas in plane samples and nanotubes located in
external magnetic fields \cite{PA-arXiv-1480-2014 2D system}. It was found that in two-dimensional electron
gas Langmuir wave replaced by the couple of hybrid waves
by considering separate spin-up electrons and spin- down electrons
evolution. One of them is the modified Langmuir wave and the other is
SEAW. Surface SEAWs was considered in Ref. \cite{Andreev 1512}. Linear interaction between surface SEAW and surface Langmuir wave (surface plasmons) is found in Ref. \cite{Andreev 1512}. In order to discuss the in-depth analysis of
SEAW, which was predicted by method of SSE-QHDs, method of separate spin evolution quantum
kinetics, which separately describes spin-up and spin-down electrons, was
developed in Ref. \cite{PA-arXiv-1409-2014 Landau damping}. By applying this
method, the effects of SSE on the real dispersion and
Landau damping of SEAW were addressed and real and
imaginary parts of spectrums of ion-acoustic waves and zeroth sound have
also been found. Nonlinear SEAWs in presence of the exchange interaction are considered in Ref. \cite{Andreev 1504}, where the existence of spin-electron acoustic soliton is demonstrated. Subsequently, in the Ref. \cite{P. A. arxiv- Oct-2015} it
has been demonstrated that the existence of SEAW
leads to an explanation of the mechanism of the electron Cooper pair formation in the
high temperature superconductors as result of electron-spelnon
interaction (spelnon is the quanta of the SEAW). Moreover, SSE-QHD model applied to study the oblique
propagation of longitudinal waves in magnetized spin-1/2 plasmas and found
that instead of two well known waves (Langmuir wave and Trivelpiece--Gould
wave) four wave solutions appeared in separate spin-up and spin-down
degenerate magnetized plasma \cite{Andreev AoP 15}.

We deal with e-p plasmas and e-p-i plasmas located in an external magnetic field $\textbf{B}_{ext}=B_{0}\textbf{e}_{z}$ and study the dispersion of waves in these systems. We assume that electrons, positrons, and ions have non-zero uniform equilibrium concentrations. The equilibrium velocity fields of all species are equal to zero. For e-p plasmas concentrations of all electrons and all positrons are equal $n_{0e}=n_{0p}$. Their spin polarization are equal to each other as well $\eta_{e}=\eta_{p}$, where $\eta_{a}=3\mu_{B}B_{0}/2\varepsilon_{Fa}$, with the Bohr magneton $\mu_{B}$ and Fermi energy $\varepsilon_{Fa}=(3\pi^{2}n_{0a})^{\frac{2}{3}}\hbar^{2}/2m$ of species $a$. Thus, we have $n_{0eu}=n_{0pd}$ and $n_{0ed}=n_{0pu}$, where $n_{0eu}=n_{0e}(1-\eta_{e})/2$, $n_{0ed}=n_{0e}(1+\eta_{e})/2$, $n_{0pu}=n_{0p}(1+\eta_{p})/2$, $n_{0pd}=n_{0p}(1-\eta_{p})/2$.
For the e-p-i plasmas we have $n_{0e}=n_{0p}+n_{0i}$. Consequently, the equilibrium concentrations of electrons and positrons are not equal each other. Therefore, their spin polarization $\eta_{a}\sim n_{0a}^{-2/3}$ are not equal each other either. Hence, parameters $n_{0eu}$, $n_{0ed}$, $n_{0pu}$ and $n_{0pd}$ are four independent parameters (one can use another set of parameters $n_{0e}$, $n_{0p}$, $\eta_{e}$, $\eta_{p}$). Relations between parameters depend on the external magnetic field.
Next, we consider linear evolution of perturbations. For the oblique propagating plane waves $\textbf{k}=\{k_{x}, 0, k_{z}\}$ we find the following dispersion equation:
$$\sum_{a=e,p}\biggl(\frac{\sin^{2}\theta}{\omega^{2}-\Omega^{2}}+\frac{\cos^{2}\theta}{\omega^{2}}\biggr)
\Biggl[\frac{\omega_{Lau}^{2}}{1-(\frac{\sin^{2}\theta}{\omega^{2}-\Omega^{2}} +\frac{\cos^{2}\theta}{\omega^{2}})U_{au}^{2}k^{2}}$$
\begin{equation}\label{SEPAWOP Obl Longit disp eq general} +\frac{\omega_{Lad}^{2}}{1-(\frac{\sin^{2}\theta}{\omega^{2}-\Omega^{2}}+\frac{\cos^{2}\theta}{\omega^{2}})U_{ad}^{2}k^{2}}\Biggr]=1,\end{equation}
where $\omega_{Lau}^{2}=4\pi e^{2}n_{0au}/m$, and $\omega_{Lad}^{2}=4\pi e^{2}n_{0ad}/m$ are the Langmuir frequencies for the spin-up and spin-down particles of species $a$, and $\omega_{La}^{2}=\omega_{Lau}^{2}+\omega_{Lad}^{2}$,
$\Omega=eB_{0}/mc$ is the cyclotron frequency, $k^{2}=k_{x}^{2}+k_{z}^{2}$, $U_{as}^{2}=(6\pi^{2}n_{0as})^{\frac{2}{3}}\hbar^{2}/3m^{2}$, $\theta$ is the angle between direction of wave propagation $\textbf{k}$ and the external magnetic field $\textbf{B}_{0}=B_{0}\textbf{e}_{z}$.

For the electron-positron plasmas we have $n_{0e}=n_{0p}$, $n_{0eu}=n_{0pd}$, $n_{0ed}=n_{0pu}$. Hence, equation (\ref{SEPAWOP Obl Longit disp eq general}) in this regime is an equation of the fourth degree relatively $\omega^{2}$ for the oblique propagation. At the propagation parallel or perpendicular to the external magnetic field equation (\ref{SEPAWOP Obl Longit disp eq general}) simplifies to equation of the second degree.

In the regime of e-p-i plasmas all four parameters $n_{0eu}$, $n_{0ed}$, $n_{0pu}$ and $n_{0pd}$ are different from each other. As the result equation (\ref{SEPAWOP Obl Longit disp eq general}) is an equation of eight degree relatively $\omega^{2}$ for the oblique propagating waves. In the regimes of the parallel or perpendicular propagation it simplifies to the equation of fourth degree relatively $\omega^{2}$.

\section{\label{sec:level1}Numerical analysis}

In our numerical analysis we use a single value of the electron concentration $n_{0e}\equiv n_{0}=10^{27}$ cm$^{-3}$. We have different regimes of the positron concentrations. To measure the concentrations of positrons in units of the equilibrium concentration of electrons we introduce parameter $\beta=n_{0p}/n_{0e}$. For the electron-positron plasmas we have $\beta=1$. For the electron-positron-ion plasmas we have $\beta\in(0,1)$. We consider two values of $\beta$ for electron-positron-ion plasmas. They are $\beta=0.5$ and $\beta=0.1$.

Changing $\beta$ at the fixed number of electrons $n_{0e}$ we change the full concentration of light particles in the system. It changes the effective Langmuir frequency, which is the frequency of Langmuir wave at $k\rightarrow0$, $\omega_{L, eff}^{2}=4\pi e^{2}(n_{0e}+n_{0p})/m=4\pi e^{2}n_{0e}(1+\beta)/m$. It creates difference in the behavior of the Langmuir wave spectrum on different figures. It is well known result which is not affect the effects caused by the spin polarization considered in this paper.

For presentation of numerical results we use the following dimensionless variables: the dimensionless frequency
$\xi=\omega/\omega_{Le}$, the dimensionless cyclotron frequency $\Sigma=\Omega_{e}/\omega_{Le}$,  and the dimensionless wave vector $\kappa=v_{Fe}k/3\omega_{Le}$.

\subsection{Electron-positron plasmas}

We start our analysis with relatively simple case of e-p plasmas ($\beta=1$). Due to the equal concentrations of electrons and positrons they have same spin polarization. Due to difference of the sign of their electric charges the numbers of spin-up electrons and spin-down positrons are equal to each other and these subspecies moves in phase. Same picture we have for spin-down electrons and spin-up positrons.

At the propagation of waves parallel or perpendicular to the external field we find two wave solutions: the Langmuir wave and SEAW, with the properties similar to the properties of these waves in e-i plasmas described in \cite{Andreev PRE 15}. 

Considering the oblique propagation of longitudinal waves in e-p plasmas we obtain Fig. \ref{SEPAWOP F1epOb} showing four waves. This regime demonstrates existence of two SEAWs, similarly to the e-i plasmas considered in Ref. \cite{Andreev AoP 15}. Hence, Fig. \ref{SEPAWOP F1epOb} shows the Langmuir wave, the upper SEAW, the Trivelpiece–-Gould wave, and the lower SEAW correspondingly, in order of the decrease of their frequency. Relative behavior of the lower SEAW and the Trivelpiece-–Gould wave is shown in Fig. \ref{SEPAWOP F1epObSm}. We see that they do not have any overlapping. The SEAW has smaller frequencies for all physically possible wave vectors.

To summarize this subsection we report of existence of two SEAWs in the e-p plasmas. We also report increase of the Langmuir wave frequency due to spin polarization entering spectrum via the Fermi pressure.

\subsection{Electron-positron-ion plasmas}

Describing longitudinal waves in e-p-i plasmas, we start with propagation of waves parallel to the external magnetic field. Fig. \ref{SEPAWOP F2} (Fig. \ref{SEPAWOP F3}) shows results for $\beta=0.1$ ($\beta=0.5$).

In the beginning of Sect. III we have described the mechanism of shift of the Langmuir wave dispersion dependence. Same effect reveals itself in the spectrum of Trivelpiece-–Gould wave, as it is depicted in Fig. \ref{SEPAWOP F8}. First of all
we see presence of four branches in both cases. Comparing these results with the well-known results and results found in Refs.
\cite{Andreev PRE 15}, \cite{Andreev AoP 15} we make the following conclusions. The upper dispersion branch
belongs to the Langmuir wave. One of these linear branches describes the SEAW as it follows from the previous subsection and Ref. \cite{Andreev PRE 15}. One of two other branches is the PAW found in Refs. \cite{Nejoh-Aust-j-phy-1996}, \cite{Tsintsadze EPJ D 11}, which exists due to different concentrations of electrons and positrons. In addition to three earlier found wave solutions we obtain an extra solution.

In the e-p-i plasmas we find four subspecies with different spin polarizations $(1+\eta_{e})/2$, $(1-\eta_{e})/2$, $(1-\eta_{p})/2=(1+\eta_{e}\beta^{-\frac{2}{3}})/2$, and $(1+\eta_{p})/2=(1-\eta_{e}\beta^{-\frac{2}{3}})/2$ instead of two existing in e-i or e-p plasmas. Therefore, we have found reacher spectrum of the spin-electron acoustic excitations.

Since new wave arises in the e-p-i plasmas due to the account of the SSE we call it the spin-electron-positron acoustic wave (SEPAW). The spectrum changes at the change of positron number. As we mention above, it changes the full number of the light particles. However, it changes the spin polarization of positrons depending on the Fermi energy of positrons. Therefore, the change of $\beta$ affects the SEPAW.

We calculate spectrum of longitudinal waves propagating perpendicular to the external field as well. We see that it increases square of all waves on $\Omega^{2}$ as it follows from Fig. \ref{SEPAWOP F4}. The frequencies of acoustic waves tend to zero at $k\rightarrow0$ in the regime of parallel propagation. Being shifted in the regime of the perpendicular propagation we have $\omega\rightarrow\Omega$ at $k\rightarrow0$. It is different for the Langmuir wave. In the regime of parallel propagation we find $\omega^{2}\rightarrow(1+\beta)\omega_{Le}^{2}$ at $k\rightarrow0$, while for the perpendicular propagation  we obtain $\omega^{2}\rightarrow(1+\beta)\omega_{Le}^{2}+\Omega^{2}$ at $k\rightarrow0$. Therefore, the shift of frequency $\omega(k\rightarrow0)$ is smaller than $\Omega$. In our case, for $\Omega=\Sigma\omega_{Le}\approx0.1\omega_{Le}$. It is hardly visible in Fig. \ref{SEPAWOP F4}, since $\omega\simeq \omega_{Le}[\sqrt{(1+\beta)}+\Sigma^{2}/2\sqrt{(1+\beta)}]=\omega_{Le}[1.05+0.01]$.

The upper linear branch in Figs. (\ref{SEPAWOP F2}), (\ref{SEPAWOP F3}), (\ref{SEPAWOP F4}) presents the SEAW similar to wave in the e-i plasmas. The middle linear branch presents the PAWs, where spin effects increases frequency of the PAW. The lower branch is the SEPAW, which is the SEAW in the subsystem of positrons. It is located lower than SEAW in the electron subsystem due to the smallar concentration of positrons in compare with the concentration of electrons $\beta<1$.

As it is directly follows from the dispersion equation (\ref{SEPAWOP Obl Longit disp eq general}) number of the dispersion branches doubles at the oblique propagation. Thus, we have eight longitudinal waves, which reduces to four branches at the parallel and perpendicular propagations. It is well-known that in e-i plasmas at the oblique wave propagation we find Trivelpiece-–Gould and Langmuir waves. The e-i plasmas with the SSE picture is more interesting. Instead of the Langmuir wave and the SEAW one can find: the Langmuir wave, the Trivelpiece-–Gould wave, the lower SEAW and the upper SEAW. If we forget about spin separation and consider e-p-i plasmas one can find the PAW \cite{Nejoh-Aust-j-phy-1996}, \cite{Tsintsadze EPJ D 11} at the parallel or perpendicular propagation. At the transition to the oblique regime we expect to find the Langmuir wave, the Trivelpiece-–Gould wave, and two PAWs. To the best of our knowledge the second PAW existing at the oblique propagation has not been reported in literature. Hence, in this paper, we report the existence of this wave.

\begin{figure}
\includegraphics[width=8cm,angle=0]{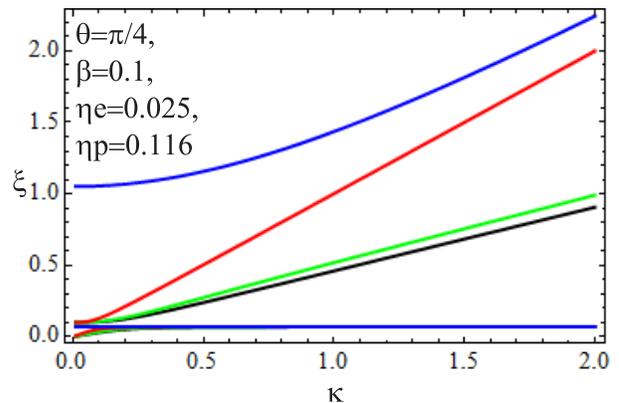}
\caption{\label{SEPAWOP F5} (Color online) The figure shows the oblique propagating longitudinal waves in the e-p-i plasmas. In this regime we find eight branches described in the text. Figure is constructed for $\beta=0.1$ and $\Sigma=0.1$.}
\end{figure}

\begin{figure}
\includegraphics[width=8cm,angle=0]{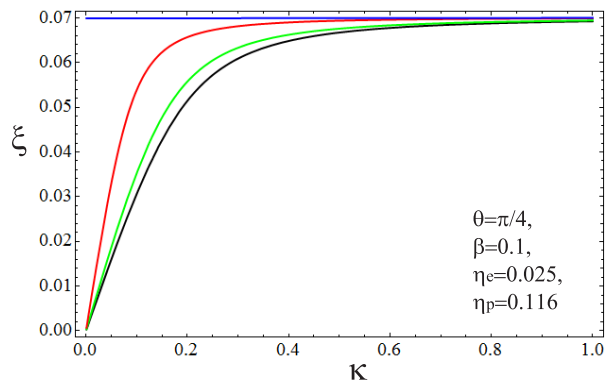}
\caption{\label{SEPAWOP F7} (Color online) The figure shows details of the low frequency part of spectrum of the oblique propagating longitudinal waves for $\beta=0.1$ and $\Sigma=0.1$.}
\end{figure}

Main subject of this paper is the e-p-i plasmas with the account of the SSE. Therefore, we obtain: the Langmuir wave, the Trivelpiece-–Gould wave, the pair of PAWs, lower and upper SEAWs, and reported for the first time, a pair of SEPAWs. The spectrum of all these waves is shown in Fig. \ref{SEPAWOP F5} for $\beta=0.1$. 
We see that the decrease of the positron concentration causes shifts of dispersion dependencies of PAWs and SEPAWs into area of larger frequencies, while the dispersion dependencies of SEAWs do not show any visible changes. Consequently the dispersion dependencies of SEAW, PAW, and SEPAW becomes closer.
It is hard to distinguish the low frequency part of the spectrum in Fig. \ref{SEPAWOP F5}. Hence, we present Fig. \ref{SEPAWOP F7}, where the low frequency part of the spectrum is depicted. Fig. \ref{SEPAWOP F7} contains the Trivelpiece-–Gould wave and lower branches of the spin-electron acoustic, positron acoustic, and spin-electron-positron acoustic waves.

\subsection{Area of applicability of obtained results}

Let us consider the following thought experiment in context of analysis of propagation of waves parallel to the external magnetic field.

Keeping a fixed number of electrons we can add a number of positrons and take away same number of ions to keep the quasi-neutrality of the system. This imaginary  quasicontinuous process we could find changes in the spectrum of the SEAW, along with the discussed above changes of the Langmuir wave spectrum. In addition to these changes we expect to find two other waves: the PAW and the SEAW. However, these two waves are predicted for plasmas with the degenerate electrons and positrons.

If we consider a temperature regime with the degenerate electrons $T\ll T_{Fe}$, and with the both subspecies of electrons are degenerate either $T\ll T_{Feu}, T_{Fed}$ (it requires relatively small spin polarization for the finite temperatures $T$), we find that positrons, for a small number of them, a non-degenerate $T\sim T_{Fp}$ or $T\gg T_{Fp}$. Here we have used the Fermi temperatures for the electrons $T_{Fe}$, spin-up electrons $T_{Feu}$, spin-down electrons $T_{Fed}$, and positrons $T_{Fp}$. In this regime, our model presented by equation (\ref{SEPAWOP cont eq electrons spin s}) and (\ref{SEPAWOP Euler eq electrons spin UP}) does not work. A strong collision damping in the system of positrons might destroy the PAW and SEPAW. Increasing number of positrons to reach conditions for the degenerate positron gas we enter area of applicability of our results.

For $n_{0e}=10^{27}$ cm$^{-3}$ and $n_{0p}=0.1 n_{0e}=10^{26}$ cm$^{-3}$, at $B_{0}=10^{10}$ G, we have $\eta_{e}=0.025$ and $\eta_{p}=0.116$. It gives $T_{Fes}(1\pm\eta_{e})^{\frac{2}{3}}T_{Fe}\approx T_{Fe}=3.47\times 10^{7}$ K and $T_{Fps}=\{0.23, 0.21\}T_{Fe}\approx 0.2 T_{Fe}=0.7\times 10^{7}$ K. Hence, our results can be used for the plasmas with temperatures below $10^{6}$ K. At larger concentrations similar results can be found for larger temperatures.

\section{\label{sec:level1}Conclusion}

Separate spin evolution in systems with the partial spin polarization has revealed itself in existence of new waves. Thereby, we have found the SEAWs in the e-p plasmas. One SEAW exists at the parallel and perpendicular propagation. Two branches exist at the oblique propagation. Their appearance is related to different Fermi pressure for the spin-up and spin-down electrons and positrons. For e-p-i plasmas we have demonstrated existence of pair of SEAWs, pair of PAWs, and pair of SEPAWs along with the Langmuir and Trivelpiece--Gould waves, at the oblique propagation. Three of them have been found for the first time: pair of SEPAWs and second (upper) PAW. These eight branches reduces to four branches at the parallel and perpendicular propagation. These branches are Langmuir wave, SEAW, PAW and SEPAW. The SEPAW has been reported for the first time.

\section{\label{sec:level1}Acknowledgements}
The work of P.A. was supported by the Russian
Foundation for Basic Research (grant no. 16-32-00886) and the Dynasty foundation. One of us Z. Iqbal gratefully acknowledge the higher education commission Pakistan for the financial support under IRSIP Award No. 1-8/HEC/HRD/2015/4027.

\end{document}